# TIME-DOMAIN APPROACH TO ENERGY EFFICIENCY: HIGH-PERFORMANCE NETWORK ELEMENTS IN DESIGN

Daniel Kharitonov  {dkh @ juniper dot net}

## ABSTRACT

*Energy efficiency is a corner stone of sustainability in data center and high-performance networking. Yet, there is a notable structural mismatch between networking element development practices and experiential equipment utilization. In particular, some aspects of energy usage (such as elasticity) routinely remains out of focus at the system level, restricting emerging technologies like IEEE 802.3az to physical and datalink layers.*

*Drawing from hands-on research and development in high-speed and grid networking, we identify a novel approach to energy efficiency in network engineering. In this paper, we demonstrate how the problem of efficient silicon design can be dissected into smaller sections based on the timescale of traffic processing. This new classification enables focused and concerted development that is tightly paired to resource and sustainability targets, which benefits devices on all network layers.*

*Keywords*— energy, green, network, routers, switches

## 1. INTRODUCTION

High-performance network devices have long avoided the boundaries and limits imposed by energy conservation. After all, 2009 marks just a decade since IEEE ratified the 802.3z standard and ubiquitous high-speed ethernet ports made energy consumption of network devices visible on the global scale.

Bandwidth availability in public and private networks has dramatically increased since the early days of networking, which has resulted in affordable data processing, broadband voice and video delivery on the global scale. Since then, the global connectivity infrastructure has moved through several generations of silicon (see Table I) that has been commensurate with progress predicted by Dennard's scaling law [2].

TABLE I.
CORE ROUTER ENERGY EFFICIENCY PROGRESS (1998-2007)

| Year | Fabrication Technology | Slot capacity (full-duplex) | Energy Efficiency* |
|------|------------------------|------------------------------|---------------------|
| 1998 | 180 nm                 | 3 Gbps                       | 66 watt/Gbps        |
| 2000 | 180 nm                 | 10 Gbps                      | 33 watt/Gbps        |
| 2002 | 130 nm                 | 40 Gbps                      | 14 watt/Gbps        |
| 2007 | 90 nm                  | 100 Gbps                     | 9 watt/Gbps         |

* source: Juniper Networks, Inc., see publication [1] for test methodology

At the present time, global ICT energy usage conforms to Khazzoom-Brookes postulate [3] and keeps increasing, which is a problem widely recognized at the national [4] and international [5] levels. Since the pace of the digital revolution cannot be slowed, we are now forced to take a new look at silicon efficiency and identify creative ways to save energy.

However, while the links between computing, resource availability and electric grids are being extensively researched and publicized [6] [7], relatively few authors focus on energy consumption in network devices. This paper analyses patterns and opportunities in network element energy consumption from the practical (user-level) and R&D (vendor) perspective, outlining existing and perspective approaches to energy conservation.

## 2. NETWORK ENERGY BASICS

It is widely recognized that modern data storage and transmission were made possible by the discipline of informational theory, established by Claude E. Shannon in his landmark article "A Mathematical Theory of Communication" [8] in 1948. As a practical consequence of his work, an arbitrary message can be represented through units of informational entropy. Modern computers and telecommunication devices typically use binary entropy units (bits) to store and access information, although some can represent data in more complex forms. In any case, actual information exchange happens by alternating and reading the unit states, a process that requires electronic or optical gates to transition between energy levels.

Not accounting for signal transmission (in the form of electrical current, light or radio waves) towards client (peer) devices, active energy consumption in networking equipment is primarily related to loss during the transfer of electric charges, which in turn is caused by imperfect conductors and electrical isolators. The exact rate of this consumption depends on technology (operating voltage and fabrication process), as well as the frequency of transitions and the number of gates involved. The latter is driven by the architecture and purpose of a telecom or network device.

Energy consumption in telecommunications is present in all layers of infrastructure and is not limited to datacenter switches and carrier-class routers. However, the former two device categories tend to share the "energy hog" label due to high energy use in a relatively small space, which has made them prime candidates for new technologies and energy-related improvements.

## 3. ENERGY EFFICIENCY IN TIME DOMAIN

The understanding of energy efficiency in ICT has significantly improved over the last couple of years and shifted from isolated component measures (such as PUE and DCiE), to system-level metrics like DceP [9] where efficiency is broadly defined as a ratio of energy consumption to useful workload. It is also widely accepted that in data switching and high-performance network applications, a useful workload takes the form of transmitted data, and efficiency is measured as the ratio of energy use to effective throughput [1].

However, it has further been argued that peak efficiency network metrics (such as ECR) cover only the highest end of the usability band, where network devices are fully loaded and system utilization is close to maximum. At the same time, empirical data coming from service providers suggest that approximately 50 percent of the time network equipment is utilized at half-capacity, and stands idle 25 percent of the time [10]. While network traces are typically non-stationary and vary between locations, it is quite clear that in order to understand energy consumption in network devices, we need to look at actual traffic patterns.

During the late 1990s a breakthrough in network traffic analysis occured, when many local-area and wide-area traces were shown to be statistically self-similar and have long-range dependencies [11]. One practical implication of this discovery was the fact that real networks were found to exhibit the burstiness phenomenon on wide range of time scales and applications. For example, in Fig. 1, link utilization is averaged at a mere 1.0 percent over a thousand-second order interval.

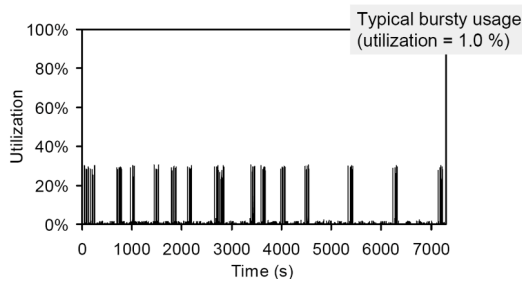

Figure 1. 100 Mbps Ethernet link utilization (source: Portland State University)

However, if we served this same link with a system sized to capacity below line rate, we would invariably end up with packet loss since bursts of self-similar traffic on the wire would have no natural length and tend to override buffers of any practical size. Also, since the superposition of several independent, heavy tailed sources yields self-similarity [12], this trend would be also observed at aggregation level (such as in packet processing engine) within a network device. This reveals that no matter what the "average" link utilization is, a high-performance network system has to be ready to operate at peak capacity at all time scales - an important constraint that we will explore in the following sections.

### 3.1. Peak energy consumption

For high-performance systems designed to operate at full capacity, the silicon packet path and all related infrastructure is sized to withstand a "hero" test of arbitrary duration – i.e. be able to forward packets at maximum speed on all ports with no "special" behavior defined for idle states. Peak energy efficiency goals are inherent to any high-performance network device design, primarily because heat dissipation is one of the major limiting factors for speed and density [13]. Despite a design focus on worst-case load conditions, most high-end network devices still show some elasticity in energy consumption due to fewer system operations and state transitions in case of reduced offered load (see Table II).

TABLE II.
ENERGY CONSUMPTION vs OFFERED LOAD

|  | 0% | 25% | 50% | 100% |
|---|---|---|---|---|
| T1600 (core router, 640Gbps full-duplex capacity), watt | 5,376 | 5,423 | 5.616 | 5,856 |
| MX960 (ethernet services router, 480Gbps full-duplex), watt | 2,925 | 3,110 | 3,209 | 3,289 |

* source: Juniper Networks, Inc., see publication [1] for test methodology

More elasticity points can be uncovered when checking energy consumption against load complexity, such as exercising key engines inversely proportional to packet length (Fig. 2).

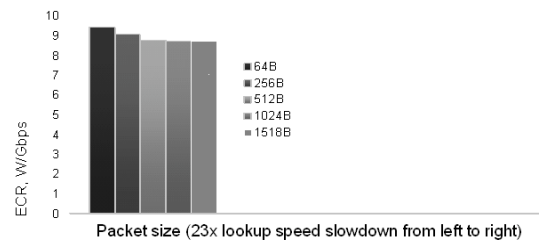

Figure 2. Energy efficiency as a function of packet size (source: Juniper Networks, T1600 router, 100% load @64x10GE ports)

However, observed changes are not very significant, which means that even best-in-class "peak efficiency" network devices will degrade in ratings at lower load levels (Fig. 3).

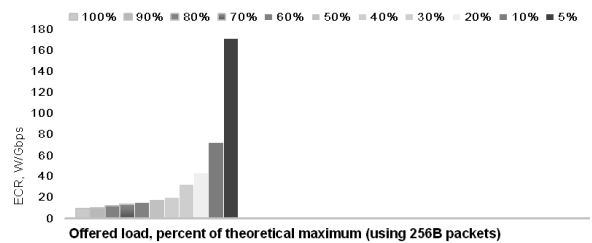

Figure 3. Energy efficiency as a function of offered load (source: Juniper Networks, T1600 router, 64x10GE ports in use)

## 3.2. Peak energy consumption summary

In the high-performance networking world, peak energy consumption goals worked well enough to keep the router/switch progress in line with demand growth and avoid the "Internet collapse" predicted by some experts [14]. "Peak energy utilization" is a fairly well-researched area, where significant progress was made before energy costs were noticeable – simply to keep the bleeding-edge routers and switches from melting down when running at full speed. However, it is often argued that the industry's ability to reduce energy consumption based on improvements in silicon lithography is about to end [15] and further improvements in energy efficiency will require non-trivial solutions.

## 3.3. Trading energy consumption for delay

The problem of poor energy efficiency at low utilization levels has long been recognized in the ICT world, but perhaps the best-known case for "energy-proportional" computing has been formulated by Luiz Barroso and Urs Hölzle of Google for IEEE Computer magazine [16]. In the article, the authors advocate for "energy-proportional" machines that consume "negligible" amounts of energy in idle states and operate in a "wide dynamic power range" characterized by tight coupling between energy consumption and utilization. Regardless of the idealistic nature of this energy-performance response, an attempt to apply identical concept to network devices will need to confront the obstacle of the real-time nature of packet processing.

Since network devices need to comply with service-level agreements (SLAs) imposed by network users, their actual ability to suspend packet processing will depend on relation between three parameters: $T_s$ (time to transition to sleep state), $T_w$ (time to transition to wake state) and $T_{sla}$. This denotes the maximum added delay or jitter the application can tolerate without violating the service contract ($T_{sla} \leq T_s+T_w$). It is important to note that the ability for a "wide range energy consumption" is orthogonal to "peak energy" efficiency. A network device with better energy elasticity may, in fact, demonstrate worse peak metrics due to additional logic to go in and out of low-power states and buffer incoming packets during transitions.

However, a hypothetical system with extensive "delay-variable energy architecture" can benefit from the same fractal nature of network traffic that made energy management so difficult in the first place: fractal models for network traffic predict the existence of extended idle states even for the busiest links [17].

The same principle of elastic energy utilization in response to delayed response currently forms the basis for the IEEE 802.3az task force (Energy Efficient Ethernet [18]).

Under 802.3az principles of operation, a network device can transition a link into a low-power state upon successful negotiation of capability with a link partner, with refresh parameters $T_q$ and $T_r$ advertised between peers and $T_w$ being optionally negotiable (Fig. 4).

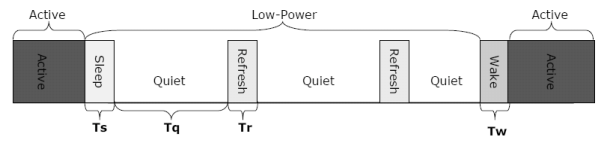

Figure 4.  IEEE 802.3az link states (source: IEEE 802.3az [19])

The main idea is that the link's PHY level can be powered down during idle periods while still retaining synchronization and being able to rapidly return back to an "active" state ($T_w \leq$ 30 usec). Changing on the order of microseconds, wait states imposed by the 802.3az draft do not affect upper OSI layers significantly and can be made compatible with most existing network devices.

To estimate the impact of a link-level energy management scheme, it's useful to take a look at the overall energy consumption within a typical high-performance network device linecard (Fig. 5)

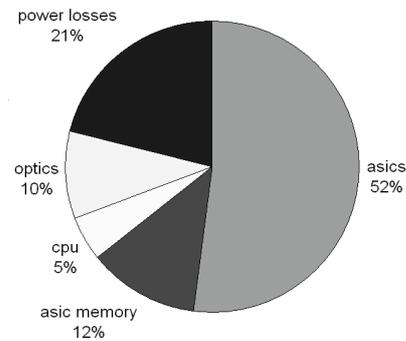

Figure 5.  Typical linecard energy usage (source: Juniper Networks)

With only ten percent of budget allocated towards link-state management, this picture suggests that high-performance network devices utilizing 802.3az protocol will not be able to change the profile of their energy consumption dramatically. Copper-based PHYs would probably yield additional improvement, although the exact amount of energy saved in a "low-power idle" state of the links can only be quantified when 802.3az support will become commercially available.

Therefore, to achieve better energy elasticity, the same principle would need to be propagated deeper into a system, preferably in the end-to-end fashion. In estimating the feasibility for improvements, the first step is to check application tolerance against system sleep/wake events, (which would be seen as jitter to the outside observer).

In general, jitter boundaries are chosen around human perception of common services, such as video (10ms acceptable jitter) and voice (30ms acceptable jitter). Network devices themselves also run protocols to ensure control plane liveness, such as BFD (sub-50ms acceptable jitter). On top of that, network equipment can be used in mission-critical (MC) applications, where any additional delay is not desired. By mapping typical linecard component wake times against applications (see Table III), we can expect serial links (including cascaded interfaces between internal linecard components) to be particularly good subjects for delay-variable energy management. Outside of a mission-critical application context, it should also be possible to integrate some energy consumption elasticity into the datapath, for example by idling lookup engines or manipulating shadow memory banks.

TABLE III
APPLICATION TOLERANCE TO SYSTEM EVENTS

|  | MC | BFD | Video | Voice |
|---|---|---|---|---|
| Pre-synchronized link into active state – 10 μs | yes | yes | yes | yes |
| Serdes bringup and frequency lock – 100 μs | yes | yes | yes | yes |
| NPU core context switching/ memory barrier – 90 ns | ? | yes | yes | yes |
| SRAM bank bringup and programming – 30ms | no | ? | ? | no |
| Embedded CPU bringup/μOS start – 2s | no | no | no | no |
| Central CPU bringup/FRU start – 100s | no | no | no | no |

### 3.4. Variable energy consumption summary

Practical implementations of "dynamic power range" are yet to find their way into the networking industry, but some of the top energy consumers, including general-purpose CPUs, ASIC memory and portions of silicon-based forwarding planes will be hard to employ in this mode of operation.

Nevertheless, considering the generally low average equipment utilization, delay-variable energy management features are potentially attractive, especially when available at low cost in mass-produced network silicon.

Research and development in this area is also hindered by the fact that - unlike peak energy management - elastic energy consumption does not necessarily result in faster and denser designs, and therefore has a limited market value (unless operational energy expenses are comparable to equipment cost). The latter condition is not unrealistic given the trend in energy costs [20], but is probably more pronounced for cost-sensitive datacenter and telecom equipment, such as low-end servers [21] and blade switches.

### 3.5. Trading energy consumption for packet loss

The inability of complex processing subsystems to change energy states quickly enough to avoid functionality loss does not necessarily eliminate the potential for energy savings. While arbitrary packet loss is generally not acceptable for high-end network equipment even at small scales, there are certain exceptions to this rule. Drawing an analogy from grid computing, wake-on-LAN feature can be deployed to dynamically configure computational clusters, while servers can maintain low-power state until summoned by an external management entity. Although a full cluster functionality during the reconfiguration phase is not available, this can be acceptable for "slow" processes with duration exceeding the wake intervals (see Table IV).

TABLE IV.
"SLOW" NETWORK PROCESS DURATION

|  | Minimum | Maximum |
|---|---|---|
| Dynamic capacity (bundle) increase | seconds | minutes |
| Planned system capacity upgrade | minutes | hours |
| Planned non-operation | minutes | no limit |
| Short-term traffic pattern (day/night) | minutes | hours |
| Long-term pattern (weekend/holiday) | days | months |

To illustrate applicability of this concept in the network devices, let's consider generic high-speed linecard building blocks (Fig. 6)

It is normal to initially deploy network platforms at a fraction of their maximum capacity in anticipation for future growth and upgrades. If a linecard carries a spare physical port adapter slot, there is a strong chance that some processing infrastructure (such as the "L2/L3 Packet Processing" block on Fig. 6) behind it is, fully operational. This is common since not every component has separate power management. It is also easier to bring up, synchronize and program all components of one board at the same time.

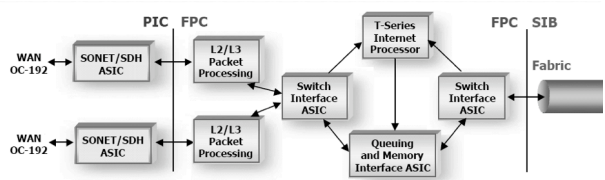

Figure 6. Typical high-performance linecard structure (source: Juniper Networks)

Another situation may happen during periodic (well-known) load reduction cycles where some performance is required, but return to full capacity can be safely delayed. During off-peak hours or days, system components like lookup engines, fabric planes and memory banks can be idled or turned off under condition that spontaneous (unplanned) load spikes will not be honored.

Idle state management operates on a different time scale from delay-variable and peak-state logic. In general, all three modes can be deployed in the same device without much overlap. To estimate the potential effect of idle state management, let's consider a router/switch equipped with with progressively less linecards. As linecards are removed, its relative efficiency drops (Fig. 7).

This occurs when fewer active ports are sharing the same common infrastructure, such as midplane, fabric, control plane and power supplies. Since planned system upgrade is a relatively slow process (see Table IV), degradation in energy efficiency can be partially offset with idle energy management at the system level. For example, additional power supplies and fabric planes can be activated in line with system capacity upgrades.

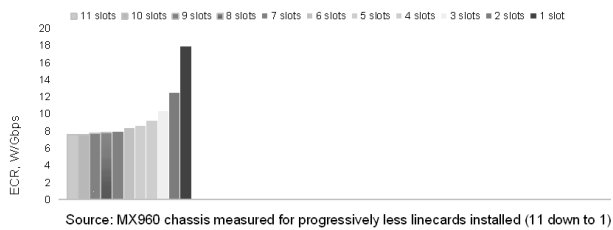

Figure 7. Energy efficiency as function of chassis fill (source: Juniper Networks, MX960 ethernet services router, 4x10GE linecards @100% utilization)

### 3.6. Idle energy management summary

Managing energy states for unused components and pluggable units may seem like a low-hanging fruit at first glance, but in reality it is quite complex because of added software and hardware required to maintain, monitor and export equipment states. The good news is that idle energy management is easily understood from a customer point of view and can be successfully marketed as a separate "energy-saving" mode. The actual usefulness of this feature heavily depends on a equipment operation profile.

### 4. CONCLUSIONS

This paper illustrates how a time-domain based classification of energy consumption modes in high-performance network devices can define the high-level map of develoment activities, risk control and expected results.

The ability to do so can prove useful not only in a strictly academic sense, but also in practical research and development, where the cost of development has to be justified against the target outcome. Fueled by the continuous growth of the global communications infrastructure, the fledgling practice of network energy management is a challenging, but exciting task.

### REFERENCES

[1] Areg Alimian. Bruce Nordman. Daniel Kharitonov. Network and Telecom Equipment - Energy and Performance Assessment. Test Procedure and Measurement Methodology

[2] Mark Bohr. A 30 Year Retrospective on Dennard's MOSFET Scaling Paper, IEEE SSCS Winter 2007, Vol. 12, No. 1

[3] Horace Herring. Does Energy Efficiency Save Energy: The Implications of accepting the Khazzoom-Brookes Postulate. EERU, The Open University

[4] U.S. Environmental Protection Agency. Report to Congress on Server and Data Center Energy Efficiency Public Law 109-431

[5] NGNs and Energy Efficiency. United Nations ITU-T Technology Watch Report #7, August 2008

[6] U.S. Energy Information Administration. Greenhouse Gases, Climate Change and Energy. DOE/EIA-X012 May 2008

[7] U.K. MetOffice. Clearer make a difference with the Facts about climate change. © Crown 2008

[8] C. E. Shannon, ``A mathematical theory of communication,'' Bell System Technical Journal, vol. 27, pp. 379-423 and 623-656, July and October, 1948, reprint available

[9] PROXY PROPOSALS FOR MEASURING DATACENTER PRODUCTIVITY © 2009 The Green Grid

[10] Verizon NEBS[TM] Compliance: Energy Efficiency Requirements for Telecommunications Equipment. Verizon Technical Purchasing Requirements VZ.TPR.9205 Issue 3, September 2008

[11] Vern Paxson and Sally Floyd, Wide-Area Traffic: The Failure of Poisson Modeling IEEE/ACM Transactions on Networking, Vol. 3 No. 3, pp. 226-244, June 1995

[12] Willinger, W., Taqqu, M.S., Sherman, R. & Wilson, D.V., "Self-Similarity Through High-Variability: Statistical Analysis of Ethernet LAN Traffic at the Source Level", IEEE Trans. Networking, 5(1)

[13] Luc Ceuppens, Alan Sardella, Daniel Kharitonov, "Power Saving Strategies and Technologies in Network Equipment Opportunities and Challenges, Risk and Rewards," SAINT, pp.381-384, International Symposium on Applications and the Internet, 2008

[14] Bob Metcalfe. The Internet is on the Verge of Collapse. Network World, 11/18/96

[15] John Shalf et al. Power, Cooling, and Energy Consumption for the Petascale and Beyond. SDSA Supercomputing 2007

[16] Luiz André Barroso and Urs Hölzle. The Case for Energy-Proportional Computing. Computer magazine. December 2007. © IEEE Inc.

[17] V.Zaborovski, Y.Podgurski and S.Yegorov, New traffic model on the base of fractional calculus, Internet http://www.neva.ru/conf/art/art8.html

[18] IEEE P802.3az Energy Efficient Ethernet Task Force

[19] Aviad Wertheimer & Robert Hays. Capabilities Negotiation Proposal for Energy-Efficient Ethernet. May 2008, Munich

[20] US EIA Short-term Energy Outlook. March 10, 2009 release.

[21] Luiz Barroso. The Price of Performance, ACM Queue, Vol. 3 No. 7 – September 2005